# Laboratory model of magnetosphere created by strong plasma perturbation with frozen-in magnetic field

**I F Shaikhislamov, Yu P Zakharov, V G Posukh, A V Melekhov, V M Antonov, E L Boyarintsev and A G Ponomarenko**

Dep. of Laser Plasma, Institute of Laser Physics SB RAS, pr. Lavrentyeva 13/3, Novosibirsk, 630090, Russia,

e-mail: ildars@ngs.ru

*Abstract:* Transient interaction of magnetic dipole with plasma flow carrying southward magnetic field is studied in laboratory experiment. The flow with transverse frozen-in field is generated by means of laser-produced plasma cross-field expansion into background plasma which fills vacuum chamber along externally applied magnetic field prior to interaction. Probe measurements showed that at realized plasma parameters effective collisionless Larmor coupling takes place resulting in formation of strong compressive perturbation which propagates in background with super-Alfvenic velocity and generates in laboratory frame transverse electric field comparable in value to expected induction one. Compression pulse with southward field after short propagation interacts with dipole and creates well defined magnetosphere. Comparison of magnetospheres created by laser-produced plasma expanding in vacuum field and in magnetized background revealed fundamental differences in structure and behavior of electric potential in plasma. In presence of frozen-in southward field direct «sub-solar» penetration of outside electric potential deep inside of magnetosphere was observed taking place with velocity close to upstream Alfven speed.

*Keywords*: magnetosphere, laser-produced plasma, magnetized background, southward field

**PACS**: 94.05.Rx, 96.25.Qr, 94.30.-d

## 1. Introduction

Southward Interplanetary Magnetic Field (IMF) frozen into Solar Wind (SW) produces a number of important and direct effects at the Earth magnetopause. These are dayside reconnection feeding Dangey cycle and substorms in the tail, flux transfer events (*Paschmann et al 1982*), increase of field-aligned currents (*Siscoe et al 2002*) and saturation of transpolar potential (*Shepherd 2007*). The change of IMF is often accompanied by pressure jump in SW which generates at the Earth magnetopause various transient effects observed at the Earth and in global numerical simulations (for example, *Moretto et al 2000, Ridley et al 2006*), such as compression and sudden impulse reaching Earth surface (*Siscoe, Formisano & Lazarus 1968*), dynamic response in ionosphere in the form of traveling vertices (*Kivelson & Southwood 1991*) and field-aligned currents (*Glassmeier & Heppner 1992*). Greatest disturbances in SW traveling as a super-Alfvenic shock are induced by Coronal Mass Ejections (CME). It is rather usual that CME contains strong magnetic field or a magnetic cloud (*Bothmer & Schwenn 1997*).

In the present paper magnetospheric processes are studied by means of laboratory models of KI-1 Facility. Experiments can supplement space observations and provide data not available by other means. We use two sources of plasma separately or in combination – induction theta-pinch and powerful $CO_2$ laser. In previous experiments on theta-pinch plasma flow interacting with magnetic dipole *(Ponomarenko et al 2001, 2008)* a detailed structure of the boundary layer and inner magnetosphere has been obtained (*Shaikhislamov et al 2012 and references therein*) based on density, magnetic field and floating potential measurements. Laser plasma experiments conducted either with (*Ponomarenko et al 2008, Zakharov et al 2014*) or without theta-pinch plasma modeling quiet SW (*Ponomarenko et al 2001, 2005, Zakharov et al 2007, 2008, 2009*) were aimed at modeling of extreme compression of the Earth's magnetosphere by super powerful CME or by artificial near-Earth releases. One of the findings was that compression of magnetosphere by pressure jump results in generation of intense field-aligned current system (*Shaikhislamov et al 2009*). Detailed measurements of the total value and local current density, of magnetic field at dipole poles and in the equatorial magnetopause, and particular features of electron motion in the current channels revealed its similarity to the Region-1 system in the Earth's magnetosphere. It was found out that magnetospheric MHD generator in low latitude boundary layer is driver of FAC in laboratory model of magnetosphere (*Shaikhislamov et al 2011, Shaikhislamov et al 2012, Shaikhislamov et al 2014*), very much in accord with model proposed earlier for the Earth (*Eastman 1976*).

However, no frozen-in transverse magnetic field as analog of IMF was present in those experiments. The aim of the given work is to realize magnetospheric interaction with a southward directed IMF. It is achieved in a novel way by launching laser-produced plasma through theta-pinch plasma in presence of external magnetic field.

The reported experiment involves several fundamental aspects. First is interaction of magnetic dipole with explosive plasma. It has been analyzed for the first time by (*Nikitin and Ponomarenko 1993*). In the MHD frame there is a single energetic parameter describing the problem – $\chi = 3W_o L_o^3 / \mu^2$ – that binds magnetic moment value $\mu$ with energy $W_o$ of spherical explosion taking place at a distance $L_o$ from the dipole center. In case of directed explosion confined in solid angle $\Delta\Omega$ a sector approximation can be used by substitute $W_o \approx 4\pi W_\Omega / \Delta\Omega$ (*Zakharov 2003, 2007*). At $\chi \ll 1$ plasma is captured as a whole while at $\chi \gg 1$ the overflowing kind of interaction takes place (*Nikitin and Ponomarenko 1995*). Experiments with laser-produced plasma in the regime $\chi \gg 1$ model extreme compression of the Earth magnetosphere by powerful CME plasma and this condition has been amply realized in the present work.

Second aspect is collisionless interaction of interpenetrating plasma flows. The specific feature of laboratory studies in this field is relatively large ion gyroradius, calculated by cross-field plasma velocity, in comparison to a scale at which significant interaction takes place. In space such conditions can be found, for example, across shocks, at reconnection sites in the Earth magnetotail, with exosphere of Mercury consisting of heavy metallic ions, or above lunar magnetic anomalies. There has been conducted many experiments in the past on laser produced plasma super-Alfvenic expansion into magnetized background plasma (for example, *Paul et al 1971, Cheung et al 1973, Tan et al 1983, Antonov et al 1983, 1985, Kacenjar et al 1986*) in which basic physics has been learned. In recent years such experiments under the trend of Laboratory Astrophysics (*Remington et al 1999, 2006*) and Laboratory Astrophysics with Lasers (*Zakharov 2003, 2013*) have been aimed to model in laboratory a colissionless shock wave. The point important to our work is that energy transfer and acceleration of magnetized background takes place at a scale of $R_* = \sqrt[3]{3N_e / 4\pi n_*}$ at which charge density of spherically expanding laser-produced plasma consisting of total number of electrons $N_e$ equals that of background plasma at density of $n_*$ (*Longmire 1963, Wright 1971*). For effective interaction plasma kinetic scales, such as ion inertia length and ion gyroradius, should be smaller than $R_*$ (*Golubev, Solov'ev and Terekhin 1978*). All this requirements have been realized in the present work and energetically effective interaction resulting in formation of a strong compressive super-Alfvenic perturbation of background was observed.

The last aspect is formation of magnetosphere by plasma flow with frozen-in magnetic field. A few experiments report on laboratory simulation with analog of southward or northward IMF. The first was a series of works by (*Podgorny, Dubinin and Potanin 1978 and ref. there in*). A picture of X-points typical of reconnection was observed upstream and downstream of magnetic dipole in case of southward IMF and at high latitude cusps for northward IMF. Measurements of slow and fast flow regimes have been made. The first was characterized by relatively small Alfven-Mach number $\leq 1.5$ and Knudsen number ~5, while in high velocity regime ion inertia length was larger than magnetosphere size thus implying strong Hall effects (*Shaikhislamov et al 2013*). Electric field was directly measured in plasma and found to correspond to given IMF and convection velocity. The specific feature of those experiments is that magnetic field was made frozen into plasma directly inside of plasma gun. However, such technique hasn't been repeated afterwards.

Other groups (*Minami and Takeya 1985, Yur et al 1999*) used instead an external magnetic field applied between plasma gun and interaction region. In (*Yur et al 1999*) the X-point was observed in the tail when southward field was applied. Study of the front region of magnetosphere (*Yur et al 2012*) revealed, that, in fact, a region of opposite southward field doesn't exists upstream of magnetopause. Thus, while authors claim that external magnetic field partially penetrates into plasma, this is not supported by presented data.

In our work we use other means of creating frozen-in magnetic field. The external magnetic field is applied along the primary plasma flow produced by theta-pinch. The field being aligned with direction of flow it isn't expelled from the plasma volume providing that magnetic pressure is stronger than thermal pressure. The perpendicular flow is generated by altogether different plasma produced by laser. It expands at Alfven-Mach number $M_A \approx 5-7$ in the volume of background plasma and launches a strong compressive perturbation. Namely this secondary flow which is generated by laser-produced plasma and which contains perpendicular frozen-in magnetic field interacts with magnetic dipole. The principal difference of employed technique is that created magnetosphere is, like laser-produced plasma, relatively short-lived. However, it suits simulation of transient effects caused by CME upon hitting Earth magnetosphere.

The main result of experiment outlined above and described in details below is that southward IMF makes a big difference on electric potential measured inside of magnetosphere. While without IMF a large potential drop exists across magnetopause created by tenuous but energetic plasma trapped inside of magnetosphere, in presence of southward IMF a process of direct penetration of outside negative potential has been observed to take place across the apex point of magnetopause.

Application of laboratory results to natural space plasma requires justification by similarity laws. Instead of a strict a physical similarity (*Podgorny and Sagdeev 1969*) proved to be most useful. Accordingly, to study a given specific process a related parameter of interest should be kept as close to natural conditions as possible, while others can be kept with degree of accuracy of order of unity, much larger or much smaller than unity. For Terrella type of experiments the most difficult to fulfill is a requirement of small ion gyroradius and inertia length in comparison to size of magnetosphere (*Schindler 1969, Baranov 1969*). The influence of ion kinetic and two-fluid effects on magnetosphere has been studied by authors in separate works (*Shaikhislamov et al 2013, Shaikhislamov et al 2014*) and present experiment was conducted at conditions when MHD rather than ion scale processes are dominant.

The paper consists of two sections on the experimental set-up and results, followed by the discussion and conclusions.

## 2. Experimental set up and results

Experiment has been carried out at KI-1 space simulation Facility, which includes chamber 500 cm in length and 120 cm in diameter operating at a base pressure of $10^{-6}$ Torr. Further on a reference frame is used with Z axis directed along the chamber symmetry axis. Induction theta-pinch fills the chamber with ionized hydrogen plasma at a density of $n_* = (2-3) \cdot 10^{13} \text{cm}^{-3}$. Flow has at velocity of $V_{*z} \approx 30 \text{km/s}$ and is sustained for duration of $30-40 \mu s$. To expand the flow immediately after theta-pinch exit aperture and to confine its propagation along the chamber a uniform magnetic field is applied. For the purpose of the present work its value was $B_{oz} = 100 \text{G}$. Due to thermal diamagnetism at electron temperature of $T_e \approx 3 \text{eV}$ theta-pinch plasma, if present, partially reduces the field value in its volume down to about $B_{*z} = 75 \text{G}$. Further on theta-pinch plasma is called a magnetized background plasma, or simply background. Alfven velocity is estimated as $V_A \approx 30 \text{km/s}$ and exceeds the ion acoustic speed of about $C_S \approx 17 \text{km/s}$. Background plasma is sufficiently collisionless as the electron-ion Coulomb collision time is much larger than electron gyroperiod $\omega_{ce} \tau_{ei} \approx 10$. In the central chamber plane background plasma encompasses a section of 70–80 cm in diameter.

Laser-produced plasma (further LP) was generated by two $CO_2$ beams of 70 ns duration and 150 J of energy each focused and overlapped into a 2 cm$^2$ spot on surface of a solid target. The target was made of polyethylene $(C_2H_4)_n$ in form of a semi-sphere with a radius of 3 cm. LP plasma consisted mostly of $H^+$ and $C^{4+}$ ions approximately in equal parts and expanded inertially in a cone with half-angle of about 30°. A bulk velocity of a main flow was about $V_o \approx 160 \text{km/s}$, while a total number of electrons per solid unit angle about $N_{e,\Omega} \approx 1.1 \cdot 10^{18}$ sr$^{-1}$ and kinetic energy $W_\Omega \approx 23$ J/sr. Due to specific pulse and tail generation mode of laser oscillator there was, besides the main plasma flow, a secondary plasma generated 550 ns later and twice as slow. At the axis of plasma expansion at a distance of $L_o$=75 cm from the target magnetic dipole was placed. Magnetic moment was oriented along Z axis and has a value of µ=5.5·10$^5$ G·cm$^3$ and a fall off time ~2.5·10$^{-4}$ sec, while the interaction lasted for about ~10$^{-5}$ sec. The dipole has an epoxy cover in form of a cylinder with a size of 5 cm. Experimental set up is shown in a photo (fig. 1) made at a time of synchronized operation of all systems. The laser target, dipole and probes have been placed in the central section of the chamber. Further on the origin of reference frame is at the dipole center with X axis pointing to laser target corresponding to the GSM frame. When necessary we use radial distance R counted from the laser target.

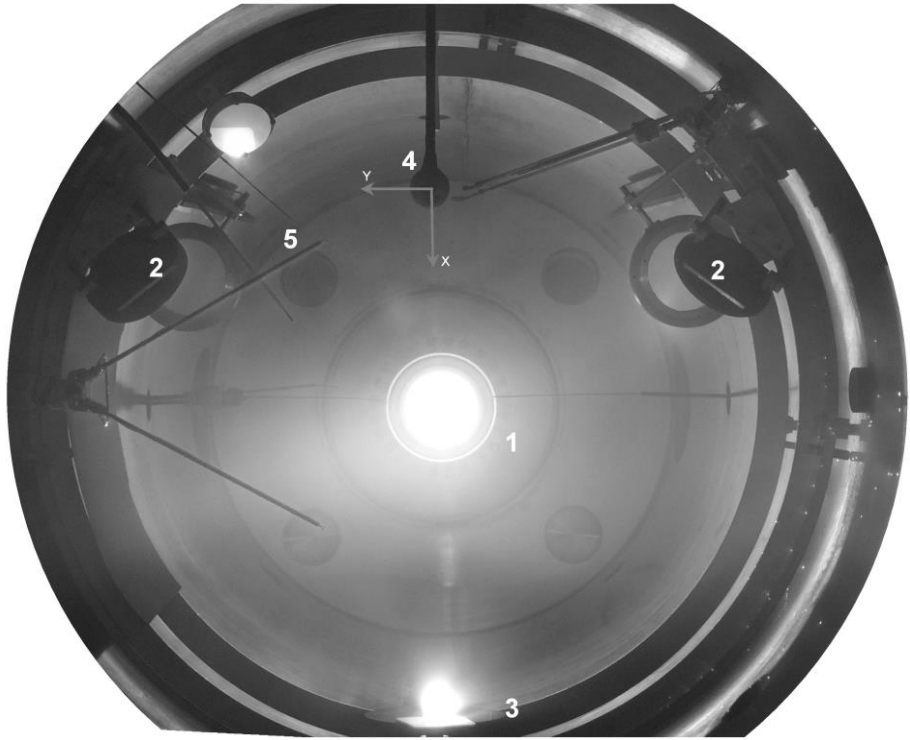

**Figure 1.** Picture of vacuum chamber and of experimental set up made with long exposure. 1 – aperture of theta-pinch; 2 – lenses and mirrors to focus and direct laser radiation; 3 – laser target; 4 – magnetic dipole; 5 – Langmuir and magnetic probes.

Diagnostics consisted of miniature electric and three-component magnetic probes with spatial resolution better than 0.5 cm which can be moved in X-Y plane. The measured values are electric probe current density $J_p$, floating plasma potential $U_p$ and magnetic field variation $\delta \mathbf{B}$. Electron density in plasma is related to directly measured Langmuir probe current as $J_p = en_e V_i \sqrt{1 + 2U_c/m_i V_i^2}$ where $U_c$ is ion collector potential, $V_i$, $m_i$ – velocity and mass of ions. For the theta-pinch hydrogen plasma with typical proton energy being much smaller than used potential $U_c = 95$ V the probe operates in a regime of density measurement $n_e (\text{cm}^{-3}) \approx 4.5 \cdot 10^{11} J_p (\text{A}/\text{cm}^2)$. For laser produced plasma typical ion energy at velocities 160-200 km/s is significantly larger than $U_c$ and probe measures electron flux while electron density is found as $n_e \approx J_p/eV_i$. To calculate pulse and energy ion composition should be taken into account. Based on previous studies we assume that LP plasma consists of protons and C$^{4+}$ ions approximately in equal parts.

Some of dimensionless parameters of the experiment in comparison to SW, CME and Earth magnetosphere are listed in table 1. Typical ram pressure of the main LP flow at the dipole location is about $p_{ram} = m_i n_i V_i^2 \approx 5 \cdot 10^3 \, dyne/cm^2$. A pressure balance estimate gives for a stand off distance $L_m = (\mu^2/2\pi p_{ram})^{1/6}$ a value of 15 cm. As will be demonstrated below, this is rather close to observed size of magnetosphere $L_m \approx 14\,cm$ produced by LP, and namely this value is used to derive dimensionless parameters. The experiment isn't aimed to model CME interaction with SW. In case of energetic super-Alfvenic magnetic cloud the properties of SW prior to CME are not very important. Described experiment is certainly relevant to a first sudden impulse phase of interaction of Earth magnetosphere with CME possessing an extremely strong southward oriented magnetic field comparable to the typical field at magnetopause, despite relatively large proton gyroradius and short duration of laboratory model. The other important condition realized in experiment is collisionless interaction of flows, as the mean free path of LP ions in background is order of magnitude larger that the distance between LP origin and dipole. The parameters of interest will be also discussed in the last section.

**Table 1.** Dimensionless parameters of experiment.

| Background plasma parameters | | Lab. | Solar Wind |
|---|---|---|---|
| Thermal beta | $8\pi n_* T_*/B_*^2$ | 0.6 | $\approx 1$ |
| Electron magnetization | $\omega_{ce}\tau_{ei}$ | $\approx 10$ | $\gg 1$ |
| | | | |
| Laser plasma parameters | | Lab. | CME |
| Mach number | $V_o/C_s$ | $\approx 10$ | $\approx 50$ |
| Alfven-Mach number | $V_o/V_A$ | $5-7$ | $\approx 50$ |
| Velocity relative to background | $V_o/V_*$ | $\approx 5$ | $\approx 5$ |
| Knudsen number | $\lambda_i/L_o$ | $\approx 100$ | $\gg 1$ ($L_o$=1 au) |
| Energetic parameter | $\chi$ | $\sim 10^4$ | $\gg 1$ |
| | | | |
| Magnetospheric parameters | | Lab. | Earth |
| Magnetopause size to dipole radius | $L_m/R_D$ | $\approx 6$ | $\sim 10$ |
| Knudsen number | $\lambda_i/L_m$ | $>5$ | $\gg 1$ |
| Reynolds number | $4\pi\sigma L_m V_o/c^2$ | $\sim 100$ | $\gg 1$ |
| Hall parameter | $L_m\omega_{pi}/c$ | $2-3$ | $\gg 1$ |
| Degree of ion magnetization | $L_m/R_L$ | $\approx 1.5$ | $\gg 1$ |
| CME duration | $\Delta t V_o/L_m$ | $\approx 5$ | $\gg 1$ |
| IMF relative intensity | $B_{IMF}L_m^3/\mu$ | $\approx 0.5$ | $\leq 1$ |

Before proceeding with a study of LP interaction with magnetic dipole immersed in magnetized background flow it is necessary to obtain a picture of plasma and magnetic field perturbation generated by dipole in otherwise uniform background flow. This is given in fig. 2 as profiles of plasma density and Z-component of magnetic field variation along the line of LP propagation. Without dipole field background parameters are more or less uniform in the region of interest – current density of about 75 A/cm$^2$ corresponding to proton density of about $n_* \approx 3 \cdot 10^{13}\,cm^{-3}$, diamagnetism level $\delta B_Z \approx -25G$ corresponding to the remaining field $B_{*_Z} \approx 75G$.

Because thermal beta is less than unity theta-pinch plasma flows mostly along magnetic field lines. The sum of dipole field with a southward (in geophysical terms) oriented uniform field creates a so called open magnetospheric configuration. A bifurcation point of zero field stands at a distance of $R_* = \sqrt[3]{\mu/B_{*_Z}} \approx 19.5\,cm$ and is marked in fig. 2 by vertical lines. Accordingly, plasma moving along field lines bypasses inner dipole region. Its density sharply decreases at the inner side of boundary $R_*$ and increases due to compression just outside of it. Magnetic field variation induced by dipole corresponds to formation of a current layer and magnetic force that balances thermal pressure gradient at the boundary. These data will be taken into account later when plotting profile of magnetosphere formed by LP.

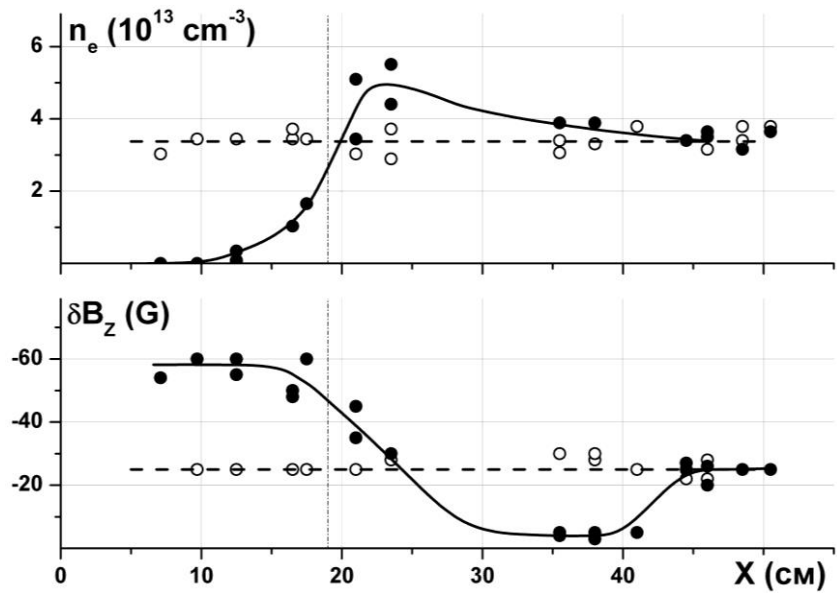

Figure 2. Profiles of plasma density (upper panel) and Z-component of magnetic field variation (lower panel) measured with (●) and without dipole field (O). Straight vertical lines mark calculated boundary of open magnetosphere.

Typical signals of Langmuir probe produced by background plasma flow, by LP expanding in vacuum and by LP expanding in background are shown in upper panel of fig. 3. As can be seen, LP consists of two pronounced flows generated by peak and tail of laser radiation. Time is counted from the moment the laser radiation arrives at the target. Before that ion current is produced only by background plasma. Total duration of LP flow $\leq 10\,\mu s$ is significantly smaller than that of background plasma flow $\approx 40\,\mu s$. Interaction of LP with background manifests itself in formation of intense density peak significantly larger than either background or LP density taken separately. The other noticeable feature is a depletion of background signal after LP flow ends (marked by arrow in fig. 3). This indicates on compression, scooping and sweeping out of background by LP.

LP generates in background strong magnetic and electric perturbations. Due to frozen-in condition LP tends totally exclude background magnetic field from its volume forming a so called cavity, while compression of background plasma at the front results in field amplification. Lower panel of fig. 3 shows a variation of the main Z-component of magnetic field and a floating plasma potential relative to a grounded wall of vacuum chamber. When the layer of compressed background arrives magnetic field sharply increases by several times. In a short while compression phase is followed by expulsion phase so during the time interval of LP flow the total magnetic field $B = B_{*z} + \delta B_z$ is practically absent (residual cavity value about 10 G). Field returns to its initial state as soon as LP action ends. The signal of probe reference electrode shows that in background plasma the potential is rather small. It rises steeply inside of compression layer and remains significant during LP flow. Note that potential is of negative value. It will be further demonstrated as well that potential generated by LP expansion in background is qualitatively different from that of LP expansion in vacuum field.

Further on we call the perturbation generated in background by LP a background pulse. Evolution of the pulse as it propagates in background is demonstrated on left panel of fig. 4. The distance from the LP origin is R=50 cm and significantly larger than in fig. 3. Ion current shows similar pronounced peak slightly broader in duration. Comparison to LP signal in vacuum field reveals that background pulse arrives at this distance later in time indicating on a deceleration. However, there is an evident overshoot preceding the background pulse. The overshoot starting point coincides with LP front in vacuum so most probably it reflects LP ions. Dynamics of floating plasma potential shows that it is generated by background perturbation and is more or less synchronous to background pulse. Magnetic field variation reveals that at this distance from LP origin there is only compression phase and practically no expulsion or cavity phase. For about $4\,\mu s$ the probe detects a layer of background plasma with frozen-in magnetic field both compressed by 2–2.5 times.

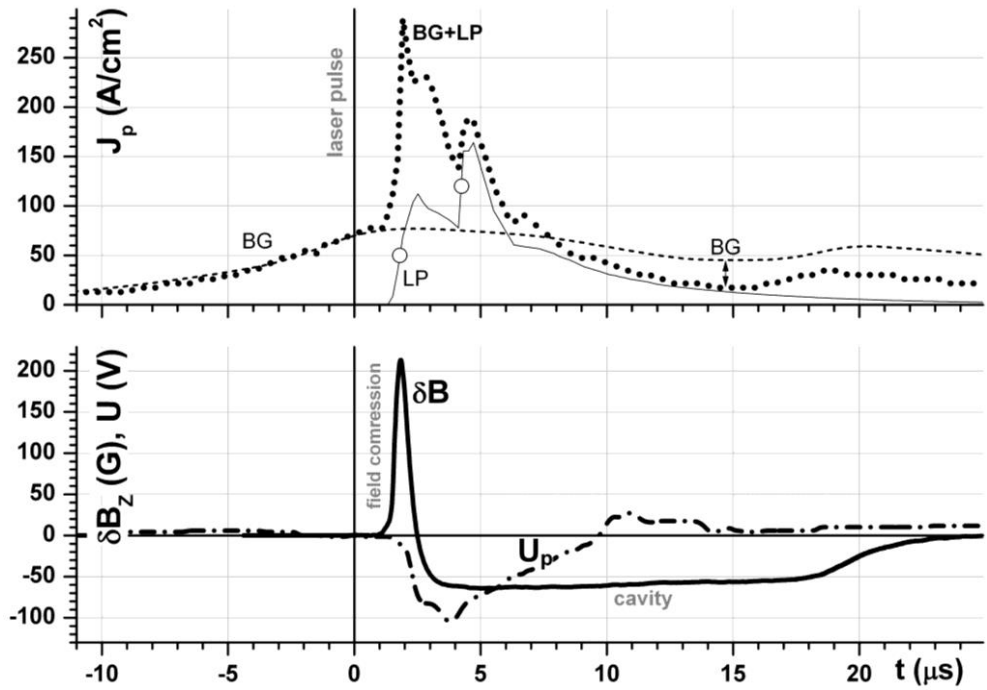

Figure 3. Dynamics of plasma flow, magnetic field and plasma potential measured at a distance of R=32 cm from the laser target (X=43 cm). Upper panel – ion current density of background plasma (BG, dashed line), of laser-produced plasma in vacuum (thin solid, circles mark positions of the first and the second front of the flow) and laser-produced plasma in presence of background (BG+LP, dotted). Laser radiation hits the target at t=0.
Lower panel – variation of Z-component of magnetic field ($\delta B$, thick solid) and floating plasma potential ($U_p$, dash-dotted).

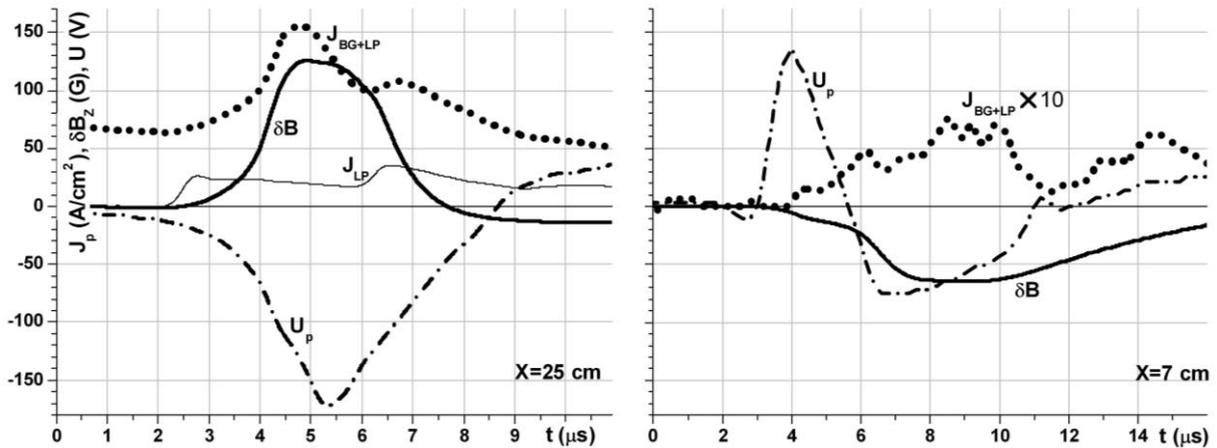

Figure 4. Dynamics of values measured outside and inside of magnetosphere at a distance of X=25 cm (left) and X=7 cm (right) from the dipole. Variation of Z-component of magnetic field (thick solid), potential (dash-dotted) and ion current density (dotted) are shown. At the left panel ion current density of laser produced plasma in vacuum ($J_{LP}$) is also shown. At the right panel ion current is multiplied by 10.

The right panel of fig. 4 presents measurements inside of the region dominated by dipole magnetic field. At such close distance dipole field effectively stops the plasma. Note that ion current density is order of magnitude smaller than at a relatively close distance but outside of magnetosphere. Direct evidence of magnetosphere formation gives the signal of magnetic field variation. It is of opposite sigh in comparison to the signal at the left panel and corresponds to compression of dipole field which is induced by magnetospheric Chapman-Ferraro current. This will be demonstrated in more details later on.

To obtain a general picture we plot time-of-flight diagrams of characteristic phase points. Left panel of fig. 5 characterizes LP motion in vacuum with and without uniform magnetic field $B_{oz}$. Fronts of the first and the second LP flow are plotted (the example is demonstrated in fig. 3 by open circles). As can be seen, velocity of the first and second front is about 220 km/s and 100 km/s respectively. From a spread of points a shot to shot variability of results can be surmised as $\pm 10\%$. The variability comes mostly from laser operation. The maximum of the first flow moves with velocity of about 166 km/s and is close to the bulk average velocity. Uniform magnetic field with strength of 100 G practically doesn't influence even the expansion of LP front up to farthest measured distance of 75 cm, all the more so the first flow maximum.

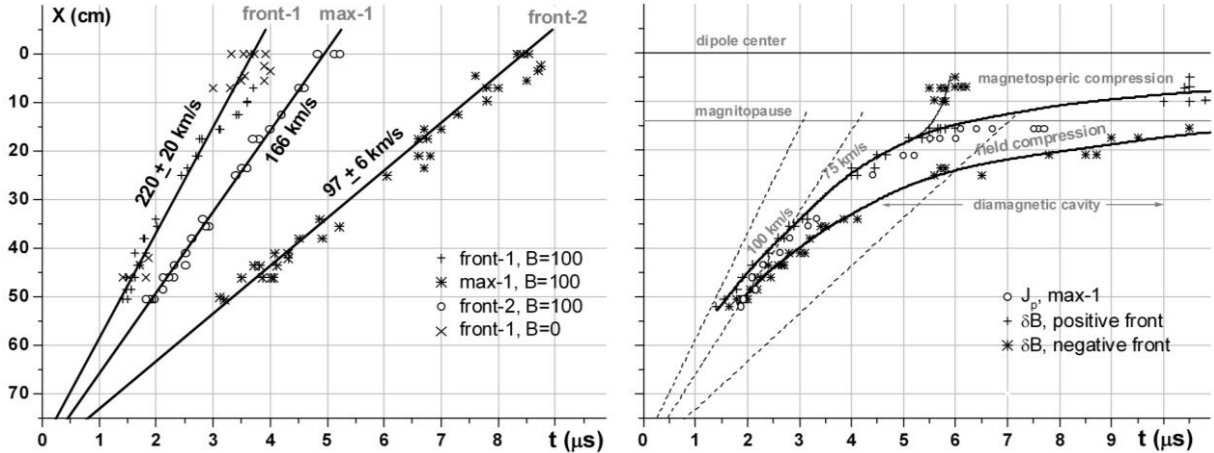

Figure 5. Time of flight diagrams. Left panel – LP free expansion in vacuum magnetic field. Points correspond to the first front (+) and first maximum (O) of ion current, as well as the second front (✱). By (✕) the points of first front obtained at zero field are shown. Straight solid lines are least square fits. Right panel – laser produced plasma expansion in magnetized background and dipole field. Points correspond to the first maximum of ion current (O), positive (+) and negative (✱) maximum of magnetic field derivative $(\partial B/\partial t)_{max}$. Dashed lines are the same as linear fits in the left panel.

Right panel of fig. 5 characterizes LP interaction with magnetized background and in presence of dipole field. First of all, the background pulse shows deceleration. While at first it appears to move practically with LP first maximum, its velocity drops to 100 km/s at a distance of about R=40 cm and to 75 km/s at about R=50 cm from the target. It is obvious that due to generation of background pulse LP first flow also experiences deceleration, especially at the front. The remnant traces of LP ions can be seen as an overshoot preceding the background pulse (see fig.4). On the other hand, velocity of the second LP flow (not shown) isn't affected by background. Note that the background pulse generated by the first LP flow always moves ahead of the second LP flow. It is possible that at later stages it is supported from further deceleration by pressure of the second LP flow.

In fig. 5 are also plotted positive and negative fronts of magnetic variation which mark the compression front and cavity front respectively (both can be clearly seen in fig. 3). Magnetic field compression exists between positive and negative fronts, while after negative front the cavity of expelled field follows. Compression front moves very close to plasma front, decelerates alongside with it and practically stops at magnetopause. Cavity front starts to slow at a distance of about R=35 cm from the target, and practically stops at R=50. Measurements with dipole switched off revealed that the compression front and background pulse are decelerated much less without dipole field. On the other hand the cavity front isn't affected by dipole field and its deceleration is caused by other reasons. Inside of magnetosphere the positive and negative fronts of magnetic perturbation switch places and constitute now a rise and fall off of magnetospheric field which is a compression of dipole field. This is clearly seen in comparison of left and right panels of fig. 4. The main point of fig. 4 and 5 is that magnetosphere is formed in interaction of dipole field with the pulse of compressed background plasma containing frozen-in compressed magnetic field and that this frozen-in magnetic field is opposite to the dipole field.

That the pulse generated in background by LP first flow is different in nature from LP itself demonstrates a plot versus distance (fig. 6) of peak electron density normalized to undisturbed background values. In vacuum field the maximum density of the first flow shows cubic fall off with the distance from plasma origin typical of LP expansion law. However, in background maximum of normalized electron density $n_e/n_*$ in the pulse decreases with distance much slower.

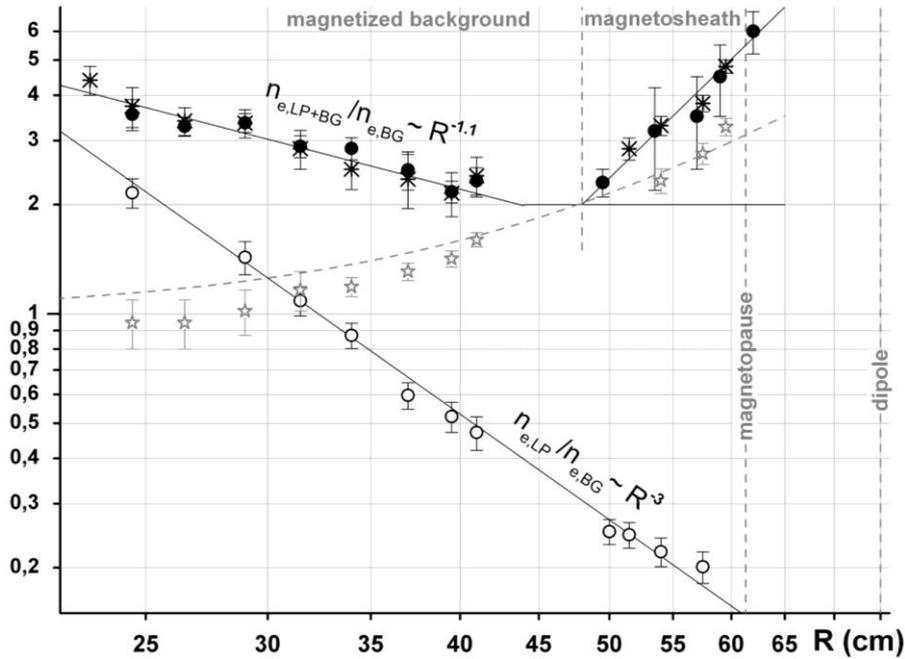

Figure 6. Maximum electron density measured by probe in dependence on distance R from the target in vacuum field (O) and in magnetized background (●). Values are normalized to electron density of theta-pinch plasma prior to LP generation. In case of vacuum field a value of $n_*=3\cdot10^{13}$ cm$^{-3}$ is taken. By (☆) data on a total number of electrons in the background pulse normalized to number of electrons in LP first flow is shown with dashed line indicating snow-plough solution. By (✱) data on relative maximum compression of magnetic field $(\delta B_z + B_{*z})/B_{*z}$ are plotted in case of LP expansion in magnetized background. By vertical lines several different regions and positions are marked.

In fig. 6 the dependence on distance of maximum magnetic compression in the pulse normalized to background field is plotted as well. One can see that the density and field compressions closely follow each other $B_z/B_{*z} \approx n_e/n_*$ indicating that their evolution is locked according to MHD equations. If the dipole is switched off the relative pulse amplitude at distances R=45–75 cm levels off at a value of $B_z/B_{*z} \approx n_e/n_* \approx 2$ (corresponding points are not shown). Dipole field, if switched on, influences the pulse at distances starting from about R=50 cm, as is clearly seen from fig 6. It courses the compression pulse to grow in relative amplitude at distances below X=25 cm from the dipole center. This is because the background pulse is stopped at magnetopause resulting in plasma accumulation. Besides the amplitude the duration of magnetic compression increases as well up to 5 μs (it can be seen in fig. 5). The region of increased plasma density and magnetic field strength in comparison to case without dipole field can be called the magnetosheath, though no shock wave has a time to develop in our experiment. Across the magnetopause plasma density sharply drops by two orders of magnitude while magnetic variation changes sign. These will be shown later when magnetosphere is described.

Besides electron density fig. 6 also shows the integral characteristic – total number of electrons in the background pulse $N_{e,\Omega} = R^2 e^{-1} \cdot \int J_p dt$ normalized to that value of LP in vacuum. Total number of electrons in vacuum doesn't depend on distance and is equal to about $N_{e,LP} \approx 1.1\cdot10^{18}$ per sr. In magnetized background approximately the same value is measured at distances up to R=30 cm, but at larger R it significantly increases. We note that this increase can't be explained by plasma deceleration at magnetopause because it doesn't change the integral number of electrons due to mass conservation.

Dashed line calculated by formula $1 + n_* R^3 / 3 N_{e,LP}$ indicates that observed rise agrees in fact with a snow-plough model.

The difference between LP flow in vacuum magnetic field and the background pulse manifests itself in electric field and plasma potential. Electric field in plasma can be derived from momentum equation for electrons:

$$\mathbf{E} = -(\mathbf{V} - \mathbf{J}/n_e e) \times \mathbf{B}/c - \nabla p_e / n_e e \qquad (1)$$

Electron inertia can be safely ignored at the scales of interest. Potential is found by solving Laplace equation $\nabla^2 U = -\text{div}\mathbf{E}$. Floating potential measured by probe differs from the plasma potential by a charging due to absorbed electron and ion fluxes, $U_p \approx U - 3.4 \cdot kT_e$. At electron temperature of about 3 eV the input of electron charging as well as electron pressure term can be ignored in comparison to values of about 100 V measured in experiment. MHD induction is described by the $\mathbf{V} \times \mathbf{B}$ term. Second term $\mathbf{J} \times \mathbf{B}/n_e e$ is a so called Hall term and can be interpreted as a deceleration force acting on plasma if it moves across field. Correspondingly, electric field generated by the Hall term is always directed inside of the LP cloud and opposite to LP expansion velocity. Associated potential inside of LP is always negative in comparison to the outside region and doesn't depend on the direction of magnetic field. On the other hand, potential due to induction term should change sign in dependence on direction of field vector.

To understand the nature of potential shown in fig. 3, 4 measurements with inversed magnetic fields, both background and dipole, have been made. First we discuss potential generated by LP in vacuum and measured at two opposite directions of magnetic fields. They are shown in upper panels of fig. 7 as spatial profiles. Data are plotted within interval of $5-6\,\mu s$, corresponding to time when magnetospheric compression and potential reach about maximum values. Left profiles are along the interaction axis X. Right profiles are along direction perpendicular both to magnetic field and velocity, that is Y axis, with minimum approach to the dipole of X=7 cm. Both profiles cross magnetosphere either frontally or laterally. Magnetopause positions, derived from reversal of magnetic variation similar to fig. 4, are marked by vertical lines. One can see that regardless of direction of magnetic field potential is always negative outside of magnetosphere and positive inside it. We note that field reversal in effect switches the sides of magnetosphere in GSM frame and some differences in observed profiles are due to general difference between west and east flanks of magnetosphere.

Negative value of potential inside of LP flow is consistent with the Hall term. Positive value of potential inside of magnetosphere is explained by penetration of small number of energetic ions due to their finite gyroradius. Penetration of magnetized electrons at the same time is much more restricted. This builds a net positive charge inside of magnetosphere to balance excess flow of positive particles. Therefore, potential associated with electric field directed to repel ions is due to the same Hall term and acts in the same way as outside of magnetosphere. This explains why the sign of potential doesn't change with field reversal.

Totally dissimilar picture is observed in case of LP expansion in background as shown in lower panels of fig. 7. Data are plotted within interval of $8-9\,\mu s$, corresponding to about median time of magnetosphere existence (see fig. 4). The main difference is that potential changes sign with field reversal everywhere. It means that in background the main term generating electric field is induction, while Hall term is relatively insignificant. It is interesting to note that prior to magnetosphere formation potential inside the region dominated by dipole field is positive like in the case of LP expansion in vacuum (see fig. 4, left panel). However, in course of magnetosphere formation it quickly becomes negative indicating on some process which levels off the difference in electric potential between inside part of magnetosphere and outer region.

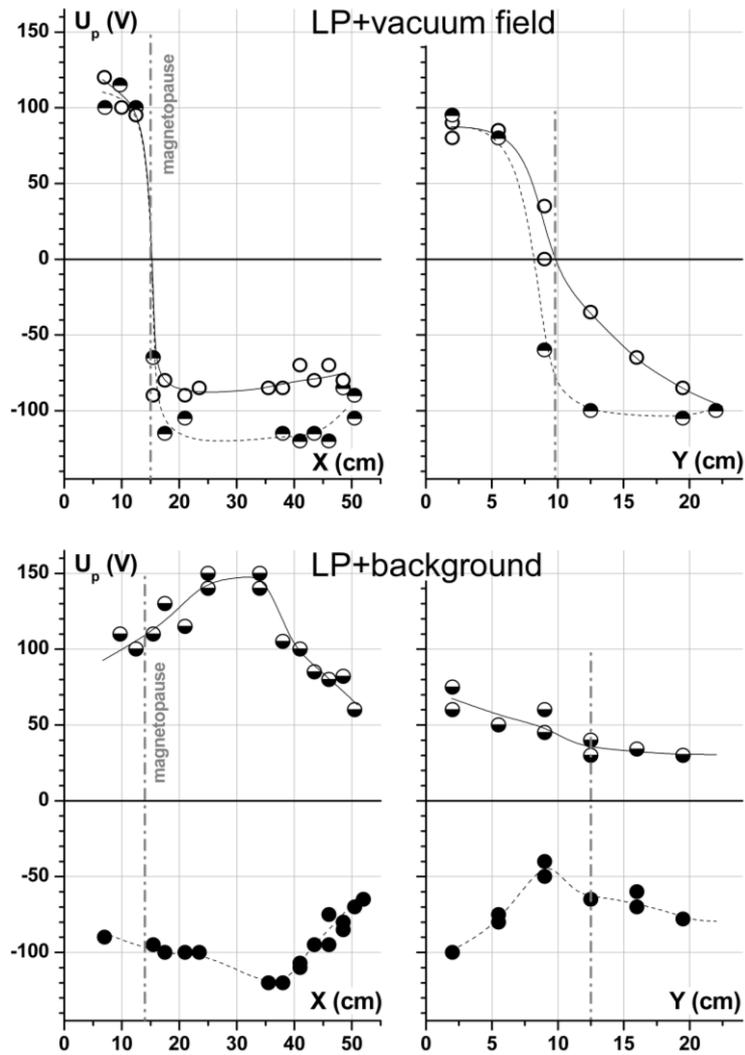

Figure 7. Profiles of floating plasma potential along X axis (left panels) and Y axis (right panels) measured at positive (O, ●) and negative (◐, ◑) value of background magnetic field and dipole moment in case of LP expansion in vacuum (upper panels) and in magnetized background (lower panels). Y–profiles are measured at a fixed minimum distance from the dipole X=7 cm.

If the electric potential is generated by cross field motion then a global Y-component of electric field should exists which, when present in SW, drives reconnection and other phenomena. The presence of such field is demonstrated in fig. 8 in profiles along Y axis which, unlike previous figure, were measured outside of magnetosphere either far off or close to magnetopause. Despite statistical spread both profiles clearly show gradient corresponding in sign to east-west induction electric field. The derived value of 2 V/cm is smaller but quite close to product of $\mathbf{V}\times\mathbf{B}/c = 2.8\,\text{V/cm}$ calculated for velocity of $V = 75\,\text{km/s}$ and field of $B_{*z} = 75\,\text{G}$. The profile taken at X=16 cm contains several points measured at reversed magnetic fields. For comparison with other points they have been transformed according to the symmetry of $\mathbf{V}\times\mathbf{B}$ term and of GSM reference frame $Y(-B) = -Y(B)$, which can be expressed as a mirror-asymmetric transformation $U(Y,-B) = -U(-Y,B)$. One can see that such points fall very close to the general trend, confirming that the observed potential is due to induction.

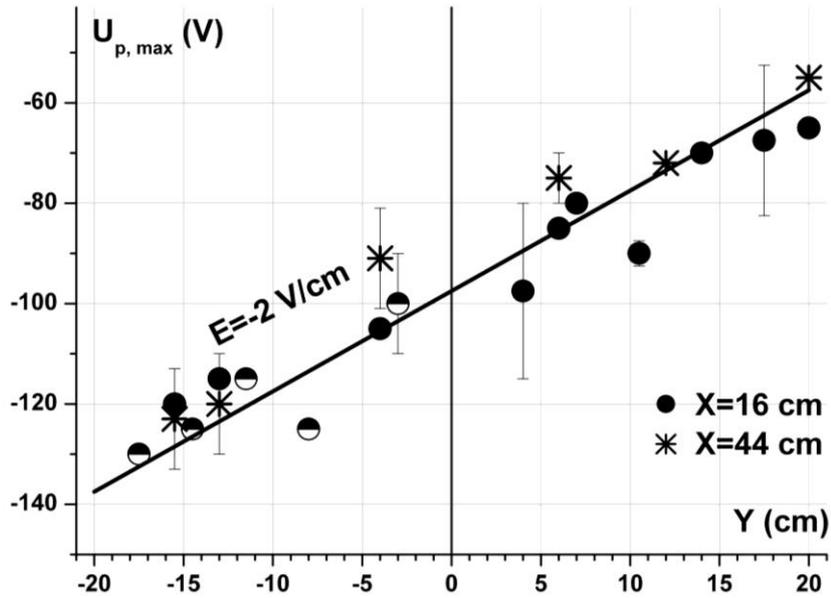

Figure 8. Profiles of maximum amplitude of floating plasma potential along Y axis measured at two minimum distances from the dipole – X=16 cm (●) and X=44 cm (✳). Data marked as ◖ have been obtained at X=16 cm and at inversed magnetic fields, and are plotted after inversion in values and Y-coordinates. Straight line gives linear approximation.

Presented so far results prove that in realized experimental conditions LP expansion in background generates strong super-Alfvenic pulse of compressed background plasma containing frozen-in compressed magnetic field. There is significant bulk velocity inside the pulse which is proved by existence of induction electric field. Next comes the main point of the paper – magnetosphere formed by plasma flow with frozen-in southward magnetic field in comparison to magnetosphere formed by LP in vacuum field. Fig. 9 complements the fig. 6 and presents profiles of magnetosphere along X axis. Upper panel shows electron density measured by probe in two cases – LP expansion in vacuum field and in magnetized background. Middle panel shows magnetic field variation $\delta B_z$, while lower panel – total field equal to $B_{tot} = B_{oz} - \mu/X^3 + \delta B_z$. In vacuum the initial field is equal to $B_{oz} = 100 G$ while in case of background a value of $B_{oz} = B_{*z} = 75 G$ is taken. Also in this case a total field variation is plotted which is a sum of variation induced by background flow around dipole prior to LP (see fig. 2) and variation due to background pulse.

Upper panel reveals that dipole forms a magnetospheric cavity in the flow. In case of magnetized background the boundary is especially sharp and plasma density outside of cavity is equal to $6 \cdot 10^{13}$ cm$^{-3}$ which is twice larger than undisturbed background density and much larger than LP density in vacuum. Middle panel reveals that dipole dominated region is divided from outside region by a thin layer of intense current. This is Chapman-Ferraro current which increases dipole field inside of magnetosphere and constitutes magnetic force that stops plasma and shields dipole. We note that the background plasma without LP makes input to magnetopause current of about 20%, so the main effect is produced by LP and background pulse. The center of current layer is positioned at about 14 cm and can be taken as magnetopause position. It is obvious that magnetospheric current generated by background pulse is several times larger than without it. Lower panel reveals that without background the total field outside of magnetosphere is very small because it is expelled by diamagnetism of LP, like it was observed in our previous experiments (*Shaikhislamov et al 2009, 2011*). However, with background present the region outside of magnetosphere contains strong magnetic field 100 G in value, opposite in direction and comparable in value to the dipole field within the current layer. The width of opposite field region is about 15 cm and comparable to the size of magnetosphere. The profiles in fig. 9 are plotted for the moment of time $8 \mu s$ and persist during $4 \mu s$ from 7 to $11 \mu s$. The lateral structure of magnetosphere (fig. 10) also shows well defined sharp magnetopause dividing inner region of compressed dipole field and reduced plasma from the outer region with opposite magnetic field.

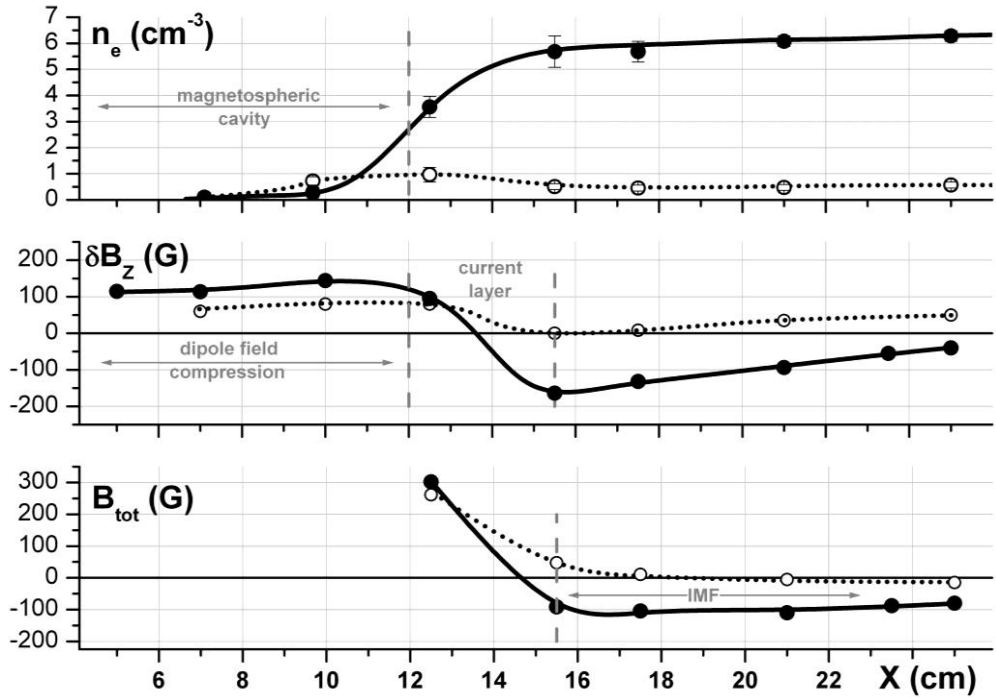

Figure 9. Profiles of electron density (upper panel), magnetic field variation (middle panel) and total field (lower panel) measured along X axis in case of LP expansion in vacuum (O, dotted lines) and in magnetized background (●, solid lines). Vertical lines mark boundaries of Chapman-Ferraro current layer.

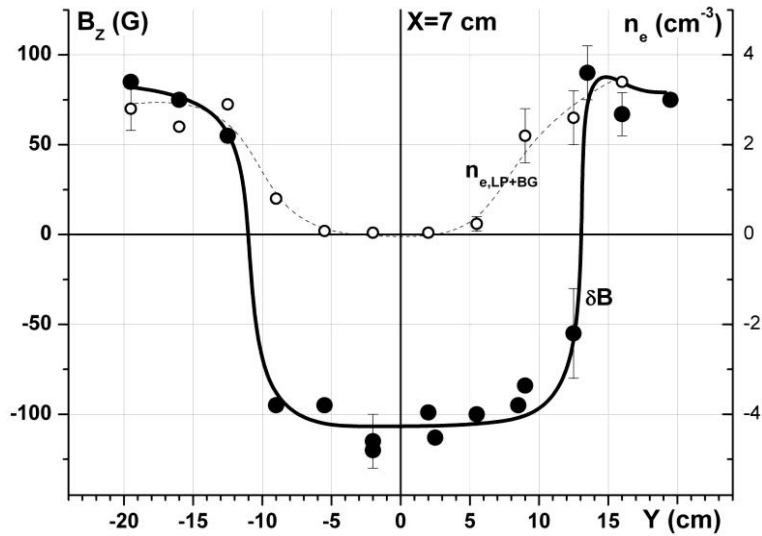

Figure 10. Lateral structure of magnetosphere. Profiles of magnetic field variation (●) and electron density (O) measured along Y axis at fixed X=7 cm are shown for the case of LP expansion in magnetized background.

## 3. Discussion and Conclusions

While detailed analysis of LP interaction with background is beyond the aim of the paper, below we give some theoretical arguments on mechanism of the background pulse generation. Estimate of Coulomb collision length of LP ions of the fist flow moving at average velocity of 160 km/s through proton plasma with density of $3 \cdot 10^{13}$ cm$^{-3}$ gives about 500 cm for protons and 8000 cm for $C^{4+}$ ions. Thus, collisions can't account for strong background sweeping and LP deceleration observed at distances of about R=25–50 cm from LP origin. The applied magnetic field can't account for LP deceleration either. Indeed, at a typical electron density $n_{e,LP} \approx 5 \cdot 10^{12}$ cm$^{-3}$ measured a distance of R=55 cm (fig. 6, 9) and bulk velocity of 160 km/s the ram plasma pressure of the first LP flow in vacuum consisting of ions with average $\overline{m}_i/\overline{z}_i = 2.6$ amounts to $p_{ram} \approx 5 \cdot 10^3$ dyne/cm$^2$ (and $\approx 2 \cdot 10^3$ dyne/cm$^2$ for the second flow). The magnetic scalar pressure at the same time is $p_{mag} = B_{oz}^2/8\pi \approx 400$ dyne/cm$^2$. Taking into account the cubic decrease of LP pressure with distance in vacuum, significant deceleration is to be expected only at a radius of about $R_B \approx 125$ cm or by 50 cm behind the dipole.

In fact, the interaction of interpenetrating flows is due to compression of magnetic field that takes place at the LP front. When LP spherically expands from the point of origin it greatly increases its volume and expels the magnetic field initially present in the space occupied later by plasma. If LP expands with super-Alfvenic velocity in presence of background plasma the expelled flux can't redistribute itself into current-free state and a thin layer of compressed magnetic field develops around LP boundary. In spherical expansion geometry such process is described by Longmire-Wright model (*Longmire 1963, Wright 1971*). It is based on free cloud expansion through undisturbed background ions with interaction being mediated by displaced electrons. The model predicts that magnetic cavity expands only as long as electron density of cloud exceeds electron density of background. It also predicts cubic fall off of magnetic field in compressed layer at the cloud front which was observed in experiments at conditions of weak interaction (*Antonov et al 1983*). When distance of equal electron densities $R_*$ exceeds ion gyroradius the ions are involved in the interaction by the so called Larmor Coupling or Magnetic Laminar Mechanism (MLM) predicted already in (*Longmire 1963*) and studied in detail by (*Golubev, Solov'ev and Terekhin 1979, Bashurin, Golubev and Terekhin 1983, Antonov et al 1985, Winske and Gary 2007*). Background ions are accelerated by transverse electric field $\mathbf{V} \times \mathbf{B}$ generated by LP in the background frame and after quarter-period of gyrorotation move in the direction of LP expansion. Maximum transverse velocity can be found from generalized momentum conservation as $V_y = z_* e B_* R/(2m_* c)$. In fact, due to flux conservation it can be calculated in the LP moving frame as a transverse velocity background ion receives in course of gyrorotation while crossing the compressed magnetic layer at the cavity front. The Larmor coupling mechanism predicts total background sweeping at distances larger than $2R_{L*}$. Theory and numerical calculations show that effective energy transfer from LP to background takes place according to value of MLM interaction parameter $\delta = R_*^2/R_L R_{L*}$ which compares radius of equal charge with gyroradiuses of LP and background ions both calculated by initial magnetic field and LP velocity (*Golubev, Solov'ev and Terekhin 1979*).

In the past a number of experiments with laser-produced plasmas has been devoted to study of super-Alfvenic collisionless interaction with magnetized background (*Paul et al 1971, Cheung et al 1973, Davis, Mahdavi and Lovberg 1976, Borovsky et al 1984, Kacenjar et al 1986*) and in recent years (*Woolsey et al 2001, Van Zeeland and Gekelman 2003, Constantin et al 2009, Niemann et al 2011, Schaeffer et al 2012, 2014, Niemann et al 2013*). Despite many attempts no substantial deceleration of LP and sweeping of background have been observed in those works, the reason being relatively large ion gyroradius, while largest interaction parameter $\delta \approx 0.3$ was achieved in experiments at KI-1 Facility (*Antonov et al 1985*). Recently order of unity value $\delta \approx 1$ has been realized due to generation of LP clouds expanding at distances of order of 100 cm in a restricted spherical cone (*Zakharov et al 2013*) thus greatly increasing density of interacting plasmas. In present experiment direct comparison of electron densities (fig. 6) gives for equal charge radius a value of $R_*$=33 cm. Based on average total number of LP electrons in the first flow in unit solid angle $N_{e,\Omega} \approx 1.1 \cdot 10^{18}$ sr$^{-1}$, the radius of cloud front at which $N_{e,\Omega}$ equals the involved number of background electrons estimates as 48 cm. The position of the fist flow maximum at the same time is equal to 36 cm. This is very close to value of $R_*$ obtained above. For the velocity of first maximum of 166 km/s the proton gyroradius in the field of $B_*$=75 G is equal to $R_{L*}$=22 cm while the gyroradius of C$^{4+}$ ions to 66 cm. Even if LP consists of only C$^{4+}$ ions the interaction parameter estimates as being very close to unity $\delta \approx 0.85$. A total sweeping of background protons is expected when first LP maximum reaches distance of R=$2R_{L*}$=44 cm.

Overview of experimental results show that observed formation of compressed background pulse is indeed due to Larmor coupling. First of all, transverse electric field accelerating background ions was directly measured (fig. 8). Second, from fig. 3 it can be calculated that gyrofrequency integral from the time of arrival of perturbation and up to the magnetic field maximum $(e/m_i c) \cdot \int B_z \cdot dt \approx 0.9$ is large enough for effective reflection of background ions by the compressed layer. When background density of electrons exceeds that of LP cloud at R ≥ 35 cm the sweeping of background ions is described by magnetic force $F_{m,x} \sim J_y B_z$. The integral of magnetic force across current layer can be approximately expressed as a potential energy of magnetic barrier $\Delta W \approx (B_{max}^2 - B_{*z}^2)/8\pi n_*$. From fig. 3 it can be seen that at the compression front total magnetic field increases from $B_{*z} \approx 75 G$ to $B_{max} \approx 275 G$ in about $0.5\,\mu s$. This yields $\Delta W \approx 58 eV$ which is enough to sweep a proton to a velocity of 108 km/s. It is close to velocity of background pulse at those distances derived from time-of-flight diagram (fig. 5).

The interesting experimental observation is that there is no positive jump of electric potential at the compression front which can be expected to correlate with magnetic barrier, as can be seen in fig. 3 and 4. This is because the Hall term in (1) $J_y B_z / n_e e$ is canceled by the convection term $V_y B_z$ so the parallel electric field $E_x$ is small inside the current layer. It means that current at compression front is conducted by ions and that magnetic force is equivalent to Lorentz force. Estimate of required current velocity $J_y / n_* e$ gives values 30–50 km/s fully compatible with maximum transverse ion velocity of 75–100 km/s.

The data and above analysis shows that in case of LP expansion in background the magnetic dipole interacts with the generated background pulse rather than LP itself. Compressed plasma with frozen-in field aligned with magnetic moment forms in front of dipole a magnetosphere and supports it for about $4\,\mu s$. Measurements reveal that at magnetopause the dipole field meets extended layer of opposite field which at the Earth drives dayside reconnection. Let's estimate a pressure balance at magnetopause. Because current layer is sufficiently thin ($\approx 3\,cm$) and dipole field is current free the magnetic pressure which balances plasma pressure can be calculated as integral over current layer $p_{mag} = (4\pi)^{-1} \int (d\delta B_z/dx) \cdot B_{tot} \cdot dx$. For vacuum case data in fig. 9 give $p_{mag} \approx 1.5 \cdot 10^3\,dyne/cm^2$. At the time interval $7-12\,\mu s$ the magnetosphere is supported by the second plasma flow which at the vicinity of magnetopause has a ram pressure of about $p_{ram} \approx 2 \cdot 10^3\,dyne/cm^2$. Thus, an experimentally measured balance between a ram and magnetic pressure at magnetopause is rather good in vacuum case. In case of LP expansion in magnetized background the magnetic pressure at magnetopause estimates as $p_{mag} \approx 4 \cdot 10^3\,dyne/cm^2$, or 2.5 times lager than in vacuum case. This is consistent with the fact that the pressure of background pulse is also significantly greater than the ram pressure of LP second flow. Just outside of magnetosheath (fig. 4, left panel) the pulse has a density of $6 \cdot 10^{13}\,cm^{-3}$ and velocity of 75 km/s which corresponds to ram pressure of $p_{ram} \approx 5.5 \cdot 10^3\,dyne/cm^2$.

The noticeable difference of magnetosphere created by background pulse is the behavior of electric potential. While in magnetosphere created by LP (or plasma flow generated by theta-pinch, *Shaikhislamov et al 2012*) a substantial potential drop exists across magnetopause, in the case of interest it rapidly equalizes between outside and inside regions of magnetosphere. It appears that potential penetrates inside of magnetosphere directly along flow direction X with velocity of about 30 km/s. This is demonstrated in fig. 11 where several dynamic signals registered at different points inside of magnetosphere are shown. Interestingly, penetration speed is close to the Alfven velocity just upstream of magnetopause $V_A = 30-40\,km/s$, velocity at which fast reconnection is assumed to proceed.

Moreover, analysis of dynamic behavior of electric potential reveals that inside of magnetosphere electric field is generally directed west-east that is opposite to the one outside of magnetosphere. This was observed in detail in earlier experiments (*Shaikhislamov et al 2011*) and is explained by the fact that plasma partially penetrating across magnetopause moves in the boundary layer tailward and along flanks of the dipole field. As the dipole field is northward and opposite to the southward frozen-in background field the induction electric field is opposite as well. The novel thing is that direct «sub-solar» penetration of potential as shown in fig. 11 leads to formation inside of magnetosphere of a compact region of reversed east-west electric field not observed in previous experiments. Partially this can be seen in the right lower panel of fig. 7.

The physics behind the fast potential penetration inside of laboratory magnetosphere is a subject of great interest in view of related phenomena observed at the Earth magnetopause and will be studied in future experiments.

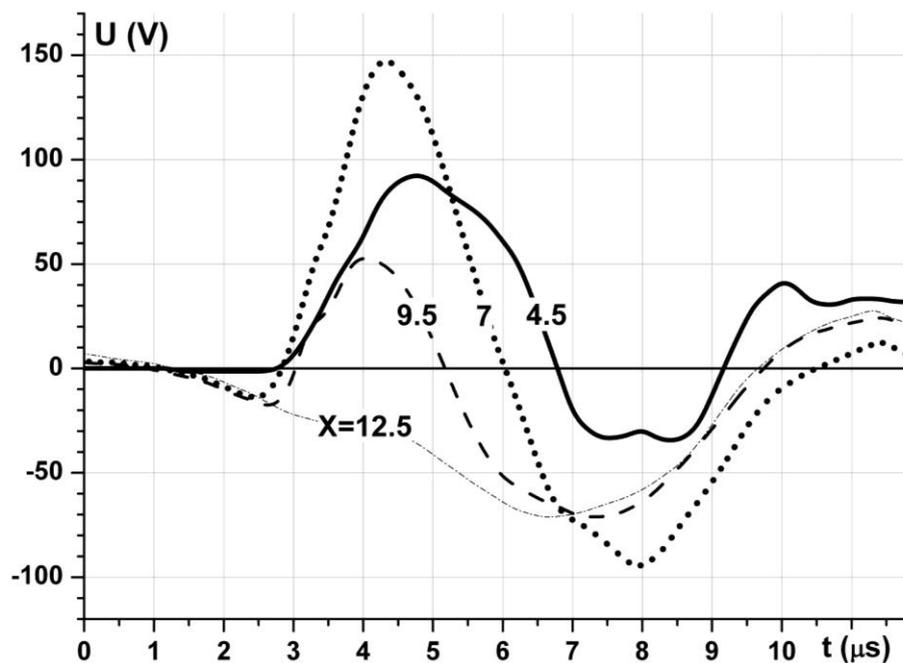
Figure 11. Dynamic of plasma floating potential obtained inside of magnetosphere at several points along X axis (marked by corresponding numbers in cm) in case of LP expansion in background.

**Acknowledgements**
This work was supported by SB RAS Research Program grant II.10.1.4 (01201374303), Russian Fund for Basic Research grant 12-02-00367, OFN RAS Research Program 15 and Presidium RAS Research Program 22.